\documentstyle[12pt,twoside]{article}

\begin{document}

\baselineskip 14pt                      

{
\renewcommand{\thefootnote}{\fnsymbol{footnote}}
\footnote[2]{\it S. Kato et al. (eds.), Basic Physics of Accretion Disks, ***--***.} 
\footnote[2]{\copyright 1995 Gordon and Breach Science Publishers.}
}

\begin{center}

{\large \bf The Physics of Black Hole X-ray Transients} \\

\vspace{0.5cm}

{J. Craig {\sc Wheeler}$^{1}$, Soon-Wook {\sc Kim }$^{1}$, 
Michael {\sc Moscoso}$^{1}$, Masaaki {\sc Kusunose}$^{1}$,
Shin {\sc Mineshige}$^{2}$} \\

{\it  1. Department of Astronomy, University of Texas, Austin TX, 78712, USA} \\
{\it  2. Department of Astronomy, Faculty of Science, Kyoto University, 
Kyoto 606-01, Japan} \\

\end{center}

\markboth
{Wheeler, Kim, Moscoso, Kusunose, Mineshige}
{The Physics of Black Hole X-ray Transients }

\vspace{0.5cm}

\begin{center}
{\bf Abstract}
\end{center}
\begin{quotation}
\noindent
The most plausible mechanism for triggering the outburst of 
black hole candidate X-ray transients is the ionization thermal instability.
The disk instability
models can give the observed mass flow in quiescence, but not the 
X-ray spectrum.  Self-irradiation of the disk in outburst may 
not lead to X-ray reprocessing as the dominant source of optical light,
but may play a role in the ``reflare."  The hard power-law spectrum
and radio bursts may be non-thermal processes driven by the flow of
pair--rich plasma from the disk at early times and due to the formation
of a pair--rich plasma corona at late times.  The repeated outbursts
suggest some sort of clock, but it is unlikely that it has anything
to do with a simple X-ray heating of the companion star.
\end{quotation}

\vspace{0.5cm}

\begin{flushleft}
{\bf 1.  Introduction}
\end{flushleft}

     Many of the X-ray novae discovered in recent years are 
black hole candidates by direct measure of their mass function
or by shared properties (McClintock and Remillard 1986, 
Tanaka {\&} Lewin 1995 and references therein).
     Among the commonly observed features are 
a primary outburst maximum,
a ``secondary reflare", 50$-$80 days after maximum, and
a ``third broad bump" in the decay, a few hundred days later.
The first two features are especially marked in the  
soft X-ray and are probably associated with the geometrically
thin, optically thick accretion disk.  The latter is associated with
the hard power-law source that may be a signature of black
hole accretion.
     Nova Per and Nova Vela revealed repeated 
``mini-outbursts" after the third,
broad bump (Callanan {\it et al.} 1995,
Della Valle, Benetti, and Wheeler 1996).
     Two ``super-luminal sources", Nova Oph 1993 (GRS 1915+105) and
Nova Sco 1994 (GRO J1655-40) display jet-like outflow in the radio
(Mirabel \& Rodriguez 1994, Harmon {\it et al.} 1994)
and radio activity is commonly associated with all outbursts
(Han and Hjellming 1992).

The power-law flux rises in Nova Muscae before the soft X-ray,
but declines rapidly so that the soft flux dominates the total power at
its maximum Miyamoto {\it et al.} 1993.  
There is a strong dip in the hard flux just before
the observations of a transient line at 480 Kev (Sunyaev {\it et al.} 1992;
Goldwurm {\it et al.} 1992). 
This feature has been associated
with a red--shifted positron annihilation line.  Although it occurs at
the rest wavelength of a Li de-excitation line (Mart\'in {\it et al.} 1994),
the lack of other
de-excitation lines and the association of the feature with the 
modulation of the hard power-law flux suggests that annihilation
is still a reasonable interpretation.  The secondary reflare is essentially
a feature of the soft flux and hence of the accretion disk.  
It occurs just as the hard power-law component reaches
a minimum, for reasons that are not understood.  At about 150
days in Nova Muscae, the disk component of the soft flux plummets
and the power, including that in soft X-rays, becomes dominated by
the power-law source at the ``third broad bump."  

\begin{flushleft}
{\bf 2. Disk Irradiation}
\end{flushleft}

The disk instability (Mineshige and Wheeler 1989) naturally
gives a rapid rise and slower decline in the soft X-ray and 
optical.  It can give an exponential decay as observed in
some sources, but the origin of this is under debate
(Cannizzo, Chen, and Livio 1996).  
The mass transfer from the companion in A0620-00 
greatly exceeds that attributed to the soft X-rays from
the inner disk (Marsh, Robinson and Wood 1994; McClintock, Horne,
and Remillard 1995).  This shows that the disks are not in steady state in 
quiescence, a principle prediction of the disk instability models.  

In principle, strong irradiation
can keep the disk ionized and prevent the disk instability.
We find that for models that approximately match the 
light curve of A0620-00 and similar sources that the irradiation 
can not be that severe
(Kim, Wheeler and Mineshige 1996a,b, and the associated
contributed paper to this conference).  
The irradiation even
at modest levels can subtly affect the disk evolution
by prolonging the disk in the intermediate, metastable
partially ionized, ``stagnation" state.  One especially
interesting manifestation of this is that some regions of the
disk can be driven from the metastable state back to the hot,
ionized state.  This causes them to be brighter intrinsically
and to intercept more irradiation.  
In our current models the associated optical
flare can match the time and amplitude of the optical component
of the reflare in A0620-00.  We do not get appreciable modulation
of the mass flow through the inner edge of the disk and so do not
get an obvious increase in the soft X-ray flux, one of the
defining characteristics of the reflare.  

The models that match the light curve of A0620-00 have the 
ignition of the thermal instability in the outer portions
of the disk.  This implies that the optical light curve
should always rise before any harder flux associated with
the increase of the mass flow in the inner portions of the
disk.  The question of whether the hard or soft X-ray flux
should increase first depends on a better understanding
of the origin of the hard flux.

These models also have the interesting property that the cooling
wave slows as the density declines, but never reaches the 
inner edge of the disk.
The inner disk thus always remains in the hot ionized state.
The mass flow rates are
$\sim10^{12}$ g s$^{-1}$,
close to those constrained by observations in quiescence
(Marsh {\it et al.} 1994; McClintock {\it et al.} 1995).  The
temperatures we derive, are, however, less than 300,000 K, whereas
the quiescent X-ray spectrum implies a temperature of about
2$\times10^6$ K.  

Van Paradijs and McClintock (1994)
have argued that, as for the LMXB, the optical flux of the black
hole candidates is due to reprocessing of X-rays.
This is not clear since the black hole systems are
transient and the flux ratio will vary in time.  
In addition, the disk instability models for
X-ray transients give an adequate amount of optical light
even with no irradiation.  
For the same orbital period, black hole disks are larger than
neutron star disks.  This mass-dependent factor alone could
make a black hole disk a factor of five to ten larger in optical
emitting area than for a neutron star.  
Van Paradijs and McClintock (1994) argue that L$_{opt}$ is proportional
to L$_x^{1/2}$.  We have constructed
steady state models with no irradiation and found that the
optical luminosity depends more steeply on L$_x$ than that
with increasing disk radius.  Furthermore,
we have computed time-dependent, non-irradiated models and found
that on the decline the locus of optical and X-ray flux is
even slightly steeper than the steady state models.  This
subject requires study in greater depth.

\begin{flushleft}
{\bf 3. The Reflare and Other Bumps}
\end{flushleft}

The ``secondary reflare" is a rather common (though not universal)
phenomenon of the black hole candidates and has never been
observed in a neutron star system. It is thus worthy of 
understanding even though it is a secondary effect compared to
the primary outburst.
 
Chen, Livio and Gehrels (1993)
discuss the possibility that 
the mass transfer rate from the companion can be modulated by irradiation
from the inner disk. 
It is not clear that a burst of mass transfer would give either
the observed optical or X-ray features.  Even a sharp burst
of added mass will
be spread by the finite viscous response of the disk
(especially when the outer parts of the disk are in the cold state) so that
any later effect in the X-rays will be delayed with respect to
the optical and very spread out in time. 
There are also questions
of whether the disk blocking invoked by Chen {\it et al.} to account for the
delay of the secondary reflare is consistent with their estimates
of mass transfer and energetics that depend on irradiating the
companion.  Similar issues arise in their model for the third broad
bump.
     
Augusteijn, Kuulkers and Shaham  (1993) 
suggest an oscillation of the light curve
in the decay in which each successive burst is a ``reflection" of
the previous burst that heats the companion and drives more mass
transfer after some time delay.
This model seems to be remarkably reminiscent of the 
``mini-outbursts" in Novae Per and Nova Vela.  Augusteijn {\it et al.} even
predicted bursts in Nova Per in August 1993 and December 1993,
(but also 21 April) as observed.  
Augusteijn {\it et al.} deserve great credit for
drawing attention to the fact that there may be some ``clock"
underlying the bursts in Nova Per and perhaps other objects, but
there are still open questions concerning their particular model.
Augusteijn {\it et al.} 
did not clearly differentiate the ``second reflare" from the
``third broad bump" as we are defining them here.  They
adjust parameters of their model to fit the second reflare of GS 2000+25
in one illustration of their model, but
then calibrate the model of Nova Per on the third broad bump in
order to ``predict" the later outbursts in that system.  
It is not at all clear
that the second reflare and the third broad bump involve similar
physics.
The models of Augusteijn {\it et al.} also do not
consider the state of the disk, especially in its
cool, quiescent, low-viscosity state, in a self-consistent way.

We do not yet have a complete understanding of the physical
mechanism of the secondary reflare (or subsequent flares).
No model yet proposed can naturally account for why the
secondary reflare seems to coincide with the 
drop in the hard X-ray flux.  Nevertheless, the 
irradiated models we have investigated show that effects in
the disk alone can give optical outbursts that may be
related to the optical flares seen.  They also give us
a new perspective from which to consider questions of
the irradiation of the companion.

     Unlike the pictures proposed by Chen {\it et al.} (1993) and
Augusteijn {\it et al.} (1993), our current models show that the
direct X-ray irradiation of either the outer disk or the 
${\rm L_1}$ point is blocked by the inner disk
throughout the decay phase prior to the secondary reflare.
     The hypothesis of the X-ray-irradiated mass transfer burst models,
that the ${\rm L_1}$ point be irradiated,
therefore, seems to be contradicted by the shadowing given
in the current models (Kim {\it et al.} 1996a,b). 

\begin{flushleft}
{\bf 4. Advection}
\end{flushleft}

Narayan, Yi, and McClintock (1996) obtain a fit to both the optical
and X-ray spectra of A0620-00 in quiescence by invoking
a hot two-temperature advective disk solution in the inner disk matched
to a steady state disk in the outer portions that provides the
optical luminosity.  The advective solution, however, is of
low efficiency and requires a mass flow rate of $4\times10^{14}$g s$^{-1}$,
much higher than the estimates based on steady state, geometrically
thin, optically thick disks by Marsh {\it et al.} (1994) and McClintock
{\it et al.} (1995) and much higher than the quiescent flow rates
we obtain in these models.  The steady-state disks appended to
the advection solutions are not consistent with the quiescent
state of the disk being modeled.  
The advection solutions require some means of
severely depleting the surface density of the inner portions
of the disk as the disk approaches quiescence.  It is difficult
to see how such a solution matches physically in terms of
the surface density and angular momentum with the outer
geometrically thin, quiescent, Keplerian disk.

\begin{flushleft}
{\bf 5. The Hard Power Law, Radio Outbursts, and Positrons}
\end{flushleft}

The hard power law component is commonly assumed to be a
Comptonized thermal spectrum.  Such a simple model can
fit some objects at some epochs, but that does not make
it unique or correct.  
Such models ignore
the obvious evidence for non-thermal particles and
magnetic fields implied by the common radio outbursts
that are frequently associated with the X-ray bursts
(Han and Hjellming 1992).
The recent super-luminal sources are only the most
extreme example.  It is most likely that the non-thermal
particles and magnetic fields arise in the disk and
hence must be incorporated into models of the hard 
power-law emission.

The soft X-ray component that probably arises in the accretion
disk peaked more slowly than the hard power law flux in
Nova Muscae.  This may mean that the inner disk was incomplete
in quiescence or the early phase of the outburst.  The radius
of the geometrically thin, optically thick disk may have
shrunk in response to increased mass flow attendant with
the disk instability in the outer disk, thus giving rise
to the delayed rise of the soft flux.

The first flare of the hard flux in systems like Nova Muscae
can be associated with a non-thermal, magnetic, pair-rich
outflow (Moscoso and Wheeler 1993).  This phase shows
QPO's, correlated radio synchrotron bursts, and at least
in Nova Muscae, the line feature that is plausibly associated
with annihilation.  If this is the annihilation line,
then it is much too narrow to represent annihilation
in the region where positrons are created and hence
implies flow of some kind.

The ``third bump," as we have defined it here, may more
closely resemble a quasi-static corona of the sort 
frequently modeled in the literature (Mineshige, Kusunose, and Matsumoto
1995).  The fact that
the disk component of the soft X-ray flux declines
as this late hard component comes in strongly suggests
that the corona is displacing the geometrically
thin disk.  With a larger effective inner radius, the
disk simply becomes too cool to emit soft X-rays.

Moscoso is constructing a model to better understand the
source of the outflow in the primary outburst.  This
model consists of an inner hot, pair-rich corona represented
by a single zone.  Above this corona, photon annihilation
will generate electron/positron pairs and associated 
annihilation.  The parallel component of the average momentum of
the photons that produce pairs is assumed to represent the
bulk outflow momentum of pairs.  The remaining momentum is
randomized to provide the 
thermal component of the pair energy.  Account will be
taken of both the isotropic and anisotropic Comptonization.
This simple model will give an estimate of the typical
flow time scales, speeds, and the optical depth so
that annihilation line profiles can be estimated.  

\begin{flushleft}
{\bf Acknowledgements}
\end{flushleft}
JCW, SKW and MK thank the organizers of the meeting for hospitality and a 
very stimulating meeting.  This research is supported in part
by NASA Grants.   

\begin{flushleft}
{\bf References}

Augusteijn, T., Kuulkers, E., \& Shaham, J. 1993, A\&A, 279, L13 

Callanan, P. J. {\it et al.} 1995, submitted to ApJ

Cannizzo, J. K., Chen, W., \& Livio, M. 1996, ApJ, in press

Chen, W., Livio, M., \& Gehrels, N. 1993, ApJ, 408, L5

Della Valle, M., Benetti, S. \& Wheeler J. C. 1996, A\&A, 
	in preparation

Goldwurm, A. {\it et al.} 1993, ApJ, 389, L79   

Han, X., \& Hjellming, R. M. 1992, ApJ, 400, 304

Harmon, B. A. {\it et al.} 1995, Nature, 374, 703

Kim, S.-W., Wheeler, J. C., \& Mineshige, S. 1995a, in preparation

Kim, S.-W., Mineshige, S., \& Wheeler, J. C. 1996b, in preparation 

Marsh, T. R., Robinson, E. L., \& Wood, J. H. 1994, MNRAS, 266, 137

Mart\'in, E. L., Rebolo, R., Casares, J., \& Charles, P. A. 1994,
ApJ, 791, 1994.

McClintock, J. E., Horne, K., \& Remillard, R. A. 1995, ApJ, 442, 358

McClintock, J. E., \& Remillard, R. A. 1986, ApJ, 308, 110

Mirabel, I. F., \& Rodriguez, L. F. 1994, Nature, 371, 46

Mineshige, S., Kusunose, M., \& Matsumoto, R. 1995, ApJ, 445, L43

Mineshige, S., \& Wheeler, J. C. 1989, ApJ, 343, 241

Miyamoto, S. {\it et al.} 1993, ApJ, 403, L39

Moscoso, M. D. and Wheeler, J. C. 1993, in Interacting Binary Stars, 
	ed. A. W. Shafter (San Francisco: ASP), 100

Sunyaev, R. {\it et al.} 1992, ApJ, 389, L75   

Tanaka, Y. and Lewin, W. H. G., 1995 in X-Ray Binaries, eds.
	W. H. G. Lewin, J. Van Paradijs, \& E. P. J. Van Den Heuvel
	(Cambridge: Cambridge University Press), 126 

Van Paradijs, J. and McClintock, J. E. 1994, A\&A, 290, 133.

\end{flushleft}

\end{document}